\documentclass[aps,prl,reprint,superscriptaddress,amsmath,amssymb,]{revtex4-1}

\usepackage{graphicx}
\begin{document}


\title{Light funneling mechanism explained by magneto-electric interference}



\author{Fabrice Pardo}
\email{fabrice.pardo@lpn.cnrs.fr}
\affiliation{CNRS -- Laboratoire de Photonique et de Nanostructures,
Route de Nozay, 91460 Marcoussis, France}

\author{Patrick Bouchon}
\affiliation{CNRS -- Laboratoire de Photonique et de Nanostructures,
Route de Nozay, 91460 Marcoussis, France}
\affiliation{Office National d'\'Etudes et de Recherches A{\'e}rospatiales,
Chemin de la Huni{\`e}re, 91761 Palaiseau, France}

\author{Riad Ha{\"i}dar}
\affiliation{Office National d'\'Etudes et de Recherches A{\'e}rospatiales,
Chemin de la Huni{\`e}re, 91761 Palaiseau, France}
\affiliation{\'Ecole Polytechnique, D\'epartement de Physique, 91128 Palaiseau, France}

\author{Jean-Luc Pelouard}
\affiliation{CNRS -- Laboratoire de Photonique et de Nanostructures,
Route de Nozay, 91460 Marcoussis, France}


\date{\today}

\begin{abstract}
We investigate the mechanisms involved in the funneling
of the optical energy into sub-wavelength grooves etched on a metallic surface.
The key phenomenon is unveiled thanks to the decomposition
of the electromagnetic field into its propagative and evanescent parts.
We unambiguously show that the funneling is not due to plasmonic waves
flowing toward the grooves,
but rather to the magneto-electric interference
of the incident wave with the evanescent field,
this field being mainly due to the resonant wave escaping from the groove.
\end{abstract}

\pacs{}

\maketitle

Plasmonics, as the science of the efficient coupling of photons with free electron
gas oscillation modes at the surface of metals, 
appears as an inescapable solution for the design and realization of 
optical nano antennas \cite{Sci-Muhlschlegel2005roa}. 
Numerous cutting edge applications are based on 
nano-antennas like biosensing \cite{NatMa-Kabashin2009pnm}, gas sensing 
\cite{NatMa-Liu2011neg}, 
photovoltaic \cite{NatMa-Atwater2010pip}
or infrared photodetection \cite{Sci-Knight2011pao} which exploit the 
intense local electromagnetic field in a confined volume
\cite{Natur-Kim2008hhg, NatMa-Schuller2010pfe, PhRvL-Novotny2007ews}.
Now, the specific matter of
total photon harvesting at the nanometric scale, i.e. 
designing an antenna able to couple all the incident optical power
with a nanoabsorber, remains challenging
\cite{NatMa-Schuller2010pfe, PhRvL-Novotny2007ews, NaPho-Tang2008nsg,
Sci-Muhlschlegel2005roa}. The natural two-step antenna sequence (collection of light,
then concentration) has been extensively studied in structures made of a metallic 
subwavelength
grating surrounding
a target \cite{OptL-Thio2001elt, JaJAP-Ishi2005snp, ApPhL-Yu2006dom, Natur-Kim2008hhg, NPho-Laux2008pps, Natur-Genet2007lit}.
The underlying mechanism involves SPP excitation (collection)
and propagation (concentration) along the grating until the coupling with the target.
Such structures, though, are designed to collect light at a specific
incidence angle, which is obviously a strong practical limitation.

In contrast, quasi-isotropic perfect transmission is obtained
through very narrow slits drilled in a metallic membrane
\cite{PhRvL-Porto1999tro, PRB-Collin2001sdi}.
This perfect transmission is successfully explained by
a localized Fabry-Perot resonance in the slits \cite{JOA-Lalanne2000omm}.
However the funneling, namely the mechanism responsible for the
{\it redirection and subsequent concentration of the whole incident energy flow, 
from the surface toward the tiny aperture of the slits},
remains unclear. Yet, a pictorial
model of the underlying physics is of a key importance for the design
of efficient nanoantennas.

\begin{figure}[b]
  \includegraphics[width=1.0\linewidth]{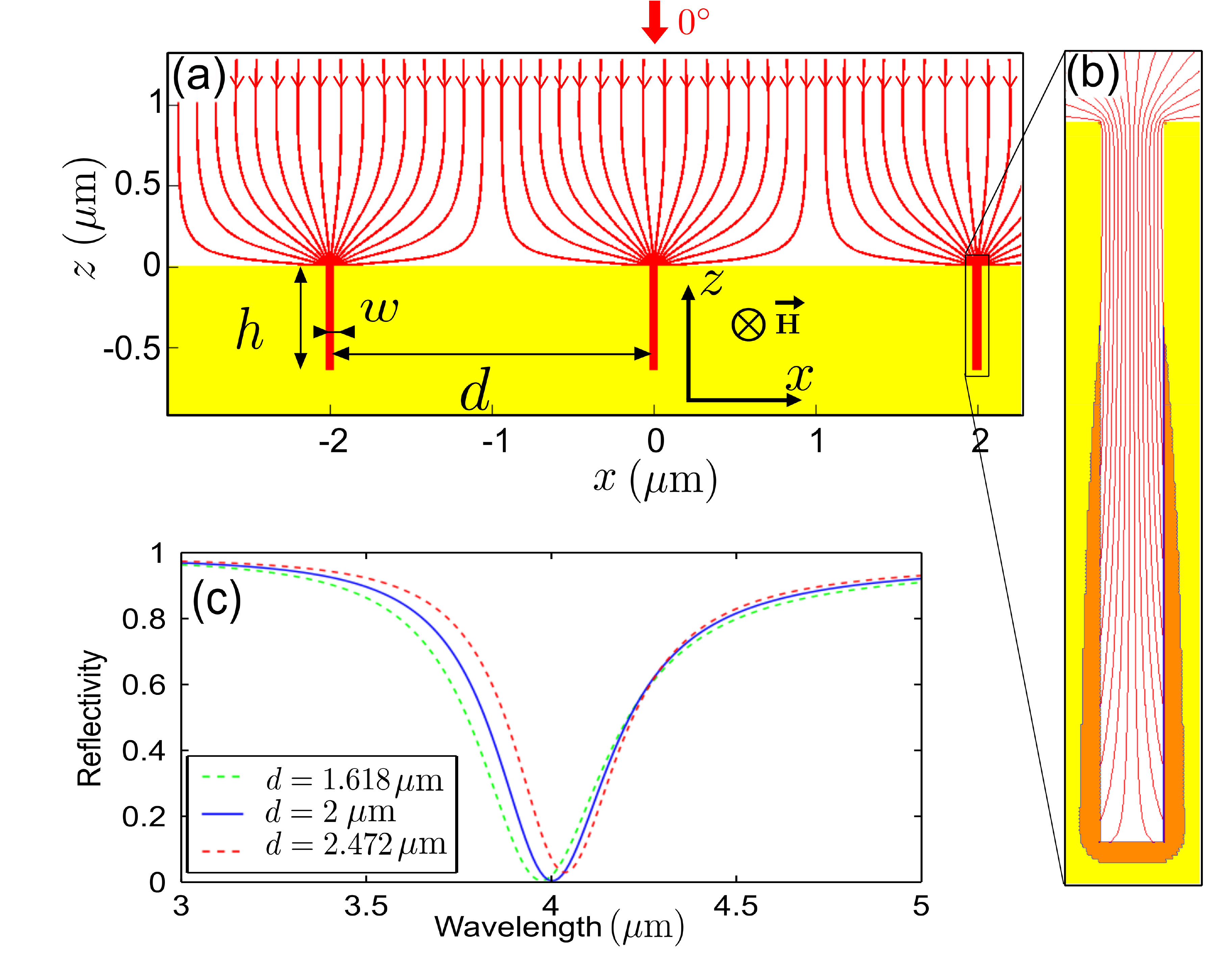}
  \caption{\label{fig:reflec}
    (a) Poynting vector streamlines are drawn on a grating of grooves of height 
    $h=640 \, \mathrm{nm}$, 
width $w=56 \, \mathrm{nm}$ and period $d=2 \, \mu\mathrm{m}$. The 
incident energy is funneled inside the grooves.
    (b) Vector streamlines inside a groove.
    The dissipation of the energy is computed
    in the metallic region, it clearly appears that this dissipation
    occurs on the 
    sidewalls of the grooves (orange volume),
    see \cite{EPAPS} for detailed field maps.
    (c) Reflectivity spectra for various values of the grating period $d=1.618 \,\mu 
    \mathrm{m}$,  $d=2 \,\mu \mathrm{m}$
and  $d=2.472 \,\mu \mathrm{m}$. The geometry of the groove is the same as before. 
The period has almost no influence 
on the resonance since it is due to the Fabry-Perot resonator
inside the grooves.
    }
\end{figure}

Such a model is given by the energetic point of view \cite{ITSTQ-Miyazaki2008hca}:
Poynting vector streamlines
distinctly show that the incident flow bends when reaching the metal surface, and
then propagates along the interface toward the slits. This fits with the
intuitive explanation, 
inspired by the SPP excitation process, that plasmonic 
waves drive the funneling
sequence \cite{OptCo-Stavrinou2002tpo}. Furthermore, quasi-cylindric waves 
were recently identified
as the dominant short-range propagation process
of the field amplitude along the surface of the grating \cite{Natur-Liu2008mto, 
PhRvL-Yang2009ccb}.
Nevertheless, even if the evanescent waves
are naturally assumed to concentrate the energy toward the apertures of the slits,
no specific study of the light funneling has so far been carried out to our 
knowledge.

In this letter, we definitely unveil the funneling process, and highlight
the unexpectedly limited role of the evanescent waves alone.
Firstly, by analysing the particular case of a groove
(for which experimental study confirm theoretical predictions \cite{EPAPS}),
we show that the 
evanescent waves
do not carry any energy through the apertures: they simply redistribute it over 
the metal surface.
Instead, we identify the {\it magneto-electric interference} (MEI)
\footnote{In order to prevent any ambiguity with the traditional
interference concept, namely $\mathbf{E}_1 \cdot \mathbf{E}_2$,
we call the term $\mathbf{E}_1 \times \mathbf{H}_2 +
\mathbf{E}_2 \times \mathbf{H}_1$
the magneto-electric interference (MEI)
of two waves 1 and 2.}
of the incident wave with the evanescent field as the main mechanism of the funneling
sequence. 
We then use a single interface analysis \cite{PhRvB-Lalanne2003paf}
to generalize our result to sub-wavelength apertures with no resonator behind.
MEI also explains the broadband extraordinary
transmission due to plasmonic Brewster angle that was
recently published by Al\`u \cite{PhRvL-Alu2011pba}
(see supplemental material \cite{EPAPS}).

Let us consider an infrared light
at $\lambda_f = 4 \, \mathrm{\mu m}$ incident onto nanometric
sized grooves periodically drilled into a gold surface.
Figure \ref{fig:reflec} (a)
shows the geometry of the grating
(width $w=56 \, \mathrm{nm}$, height $h= 640 \, \mathrm{nm}$ and
period $d$). The period is chosen so that there is no diffracted wave for all angles
of incidence (hence $d \le \lambda_f/2$). 
The light is TM-polarized
(transverse magnetic) and incident with an angle $\theta$. 
The dielectric function of gold is computed from the Drude model
$\varepsilon(\lambda) =
1 - \left[\left(\lambda_p/\lambda + i \gamma\right)\lambda_p/\lambda\right]^{-1}$
which is suited to the infrared spectral range for
$\lambda_p = 161 \, \mathrm{nm}$ and $\gamma = 0.0077$ \cite{Book-Palik1985hoo2}.
The electromagnetic analysis of this structure is done
using a B-spline method \cite{JOSAA-Bouchon2010fmm},
which can perform fast and exact computation of Maxwell equations.
The Poynting-vector streamlines show how the energy flow is funneled toward the apertures,
and dissipated mainly on the sidewalls of the grooves.
The reflectivity of the grating is plotted in Fig. \ref{fig:reflec} (c)
at normal incidence
for $d = \lambda_f/2$ and for a random value $d = 1.618 \mu \mathrm{m}$.
Although the grating is
structured on a tiny portion of its surface (less than $3\%$),
it exhibits a resonance with a total absorption at normal incidence.
We should highlight that $\lambda_f$
depends only slightly 
on the period $d$.
The absorption remains nearly total even for large incidence angles
($\theta \le 50^\circ$),
which is adapted to light collecting systems.

In order to address the funneling mechanism, we consider the electromagnetic 
field in the air,
and we split it into three terms. The magnetic field is expressed:
\begin{equation}
  H_{total}=H_{i}+H_{r}+H_{e},
  \label{fig:composantesH}
\end{equation}
where $H_{i}$ is the incoming wave,
$H_{r}$ is the reflected wave,
and $H_{e}$ is the sum of the diffracted evanescent waves.
Similar definitions can be given for the electric field components.
In the rest of this letter,
$E \times H$ stands for the mean time average value of the vectorial product
and is practically computed from complex amplitudes
as $\frac{1}{2} \Re{(\mathrm{E \times H ^{*}})}$.
Thanks to the decomposition of Eq. (\ref{fig:composantesH}),
the Poynting vector can be expressed as the sum of six terms:
\begin{equation}
  S=S_{i}+S_{ei}+S_{r}+S_{er}+S_{e}+S_{ir},
  \label{eq:S5}
\end{equation}
with
$S_i= E_{i} \times H_{i}$,
$S_{ei}= E_{e} \times H_{i} + E_{i} \times H_{e}$,
$S_r= E_{r} \times H_{r}$,
$S_{er}= E_{e} \times H_{r} + E_{r} \times H_{e}$,
$S_e= E_{e} \times H_{e}$,
and $S_{ir}=E_{i} \times H_{r} + E_{r} \times H_{i}$.
The terms $S_{i}$ and $S_{r}$
are respectively the incident and the reflected fluxes of the plane wave.
The term $S_{e}$ corresponds to the energy carried by the evanescent waves.
The term  $S_{ei}$ corresponds to the MEI
between the evanescent and the incident fields.
It is easy to prove that the six terms
of equation (\ref{eq:S5})
are flux conservative (null divergence) thus each of them can be considered
to be an independent energy flux vector in the air.
In order to simplify the discussion,
we first consider an optimized device at the resonance wavelength.
So we have no reflected wave, the fields $H_{r}$ and $E_{r}$ are null
and the Poynting vector can be expressed as
$S=S_{i}+S_{ei}+S_{e}$.

\begin{figure}[b]
  \includegraphics[width=0.45\linewidth]{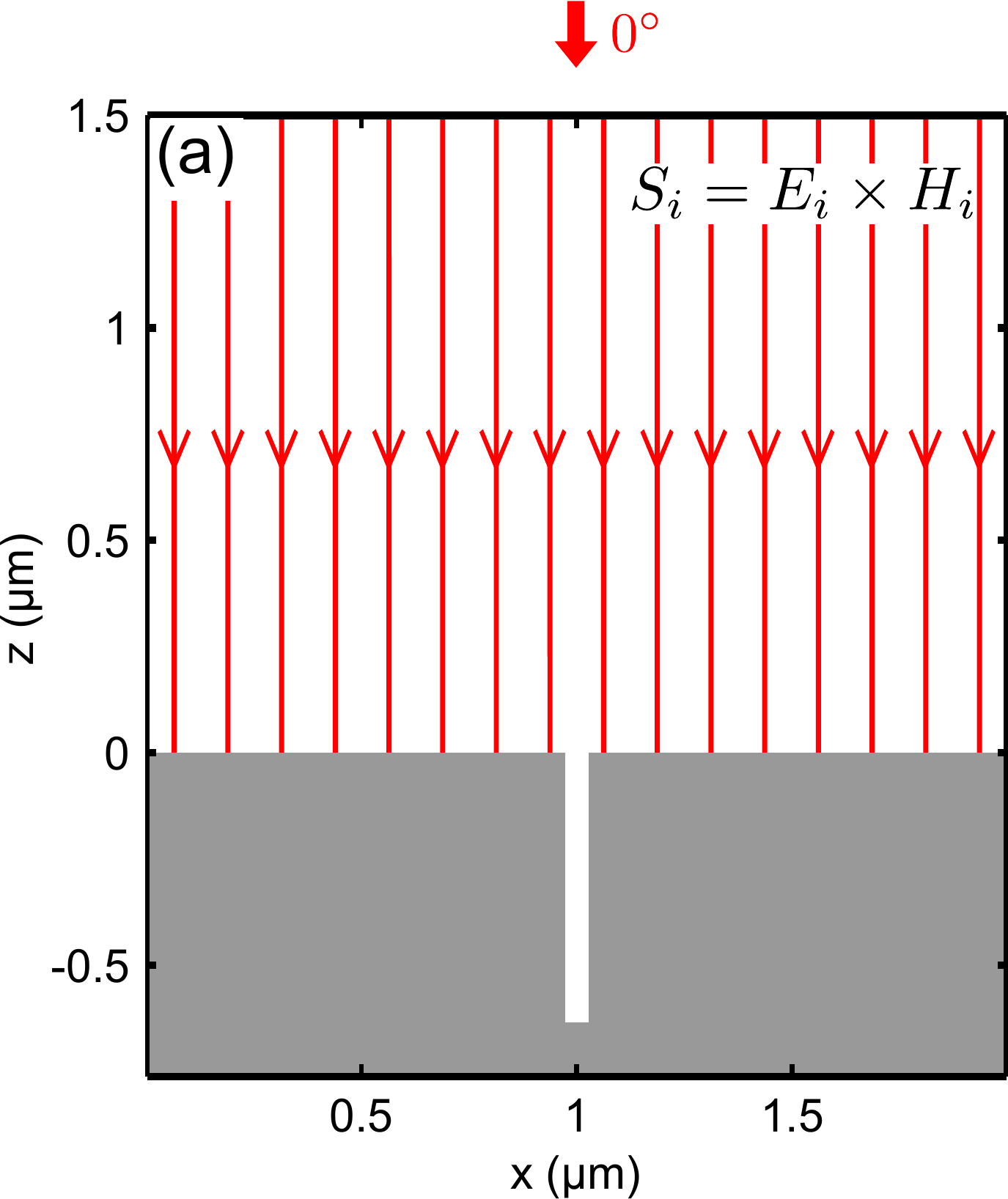}
  \includegraphics[width=0.45\linewidth]{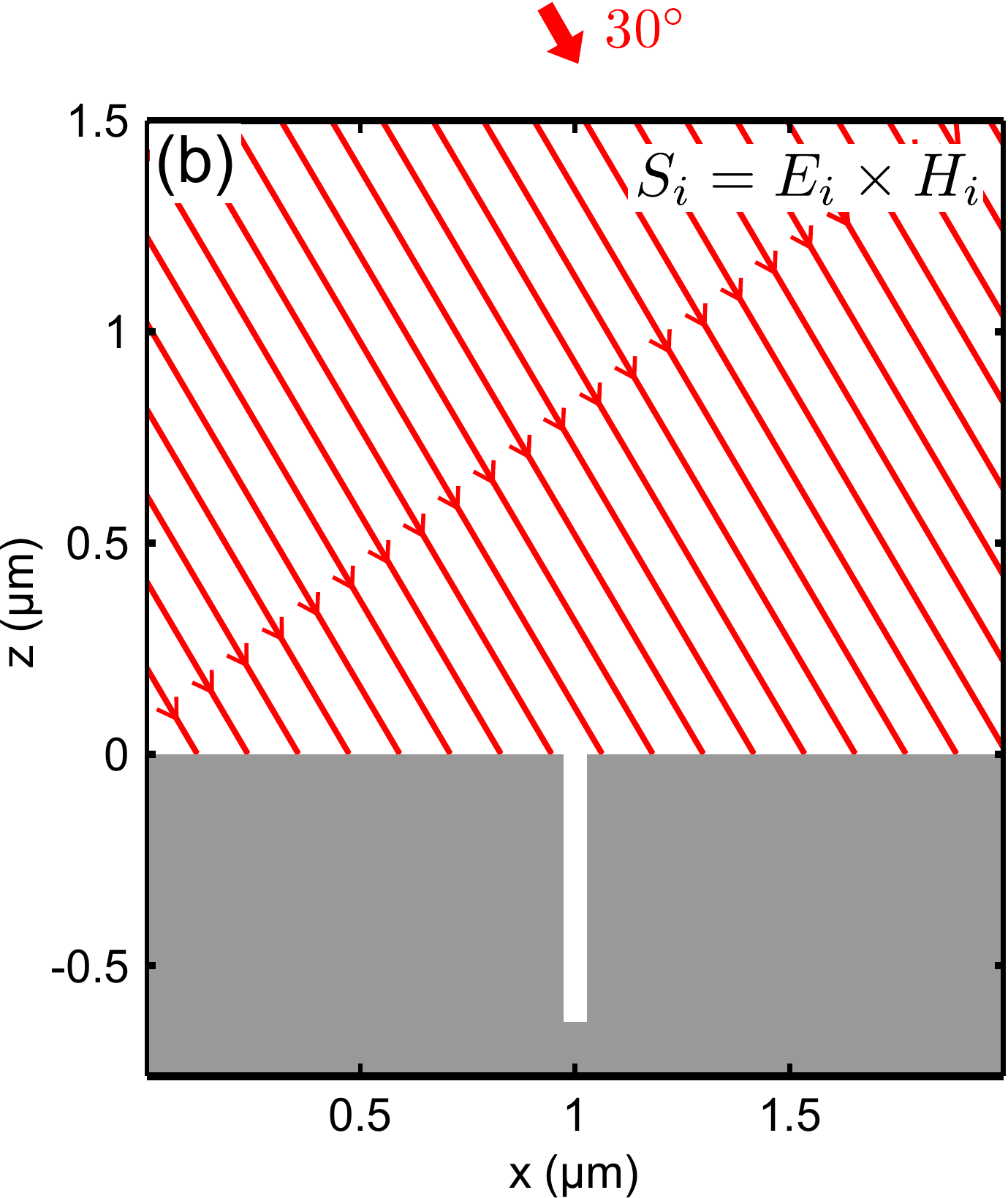}\vfill
  \includegraphics[width=0.45\linewidth]{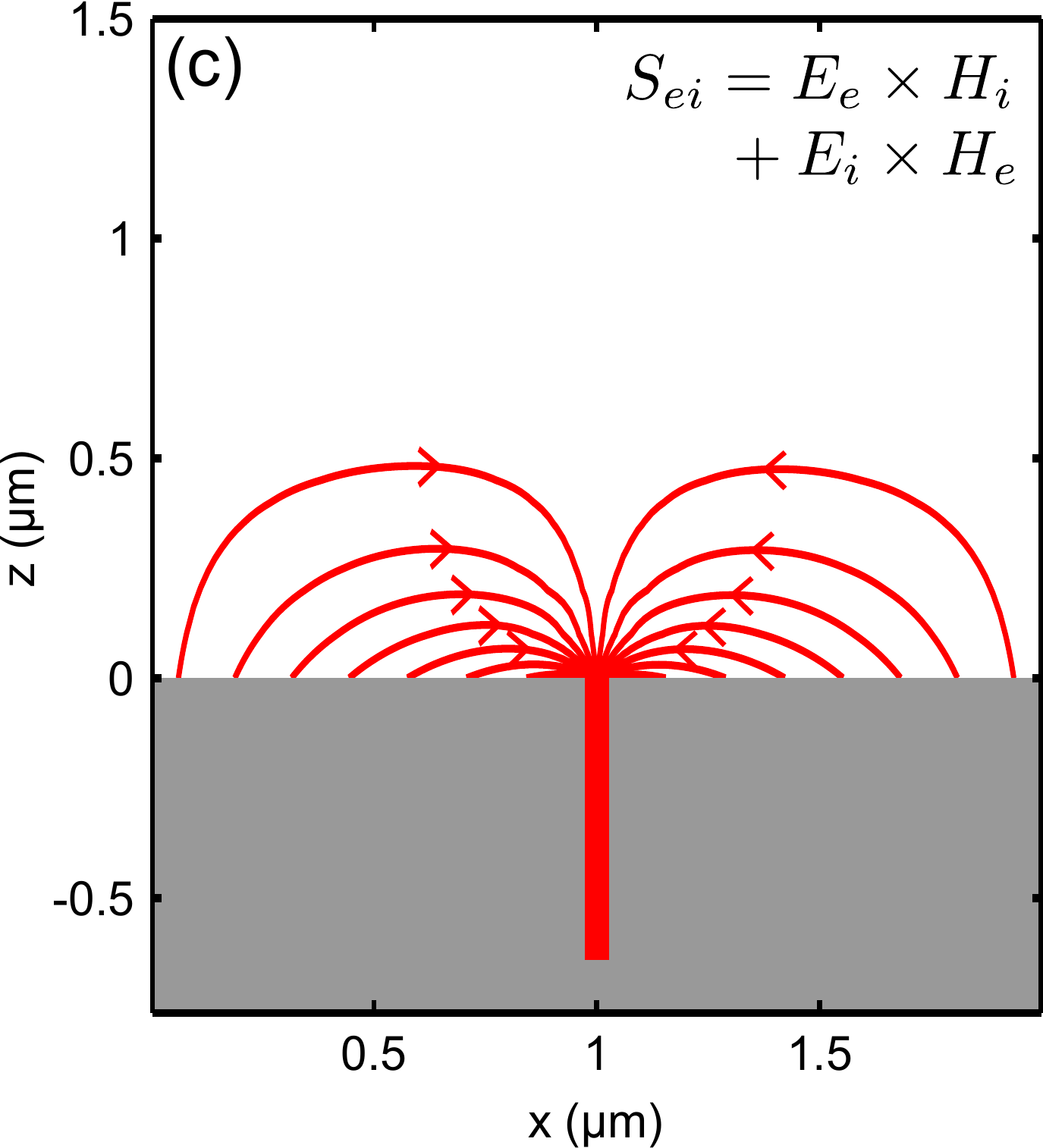}
  \includegraphics[width=0.45\linewidth]{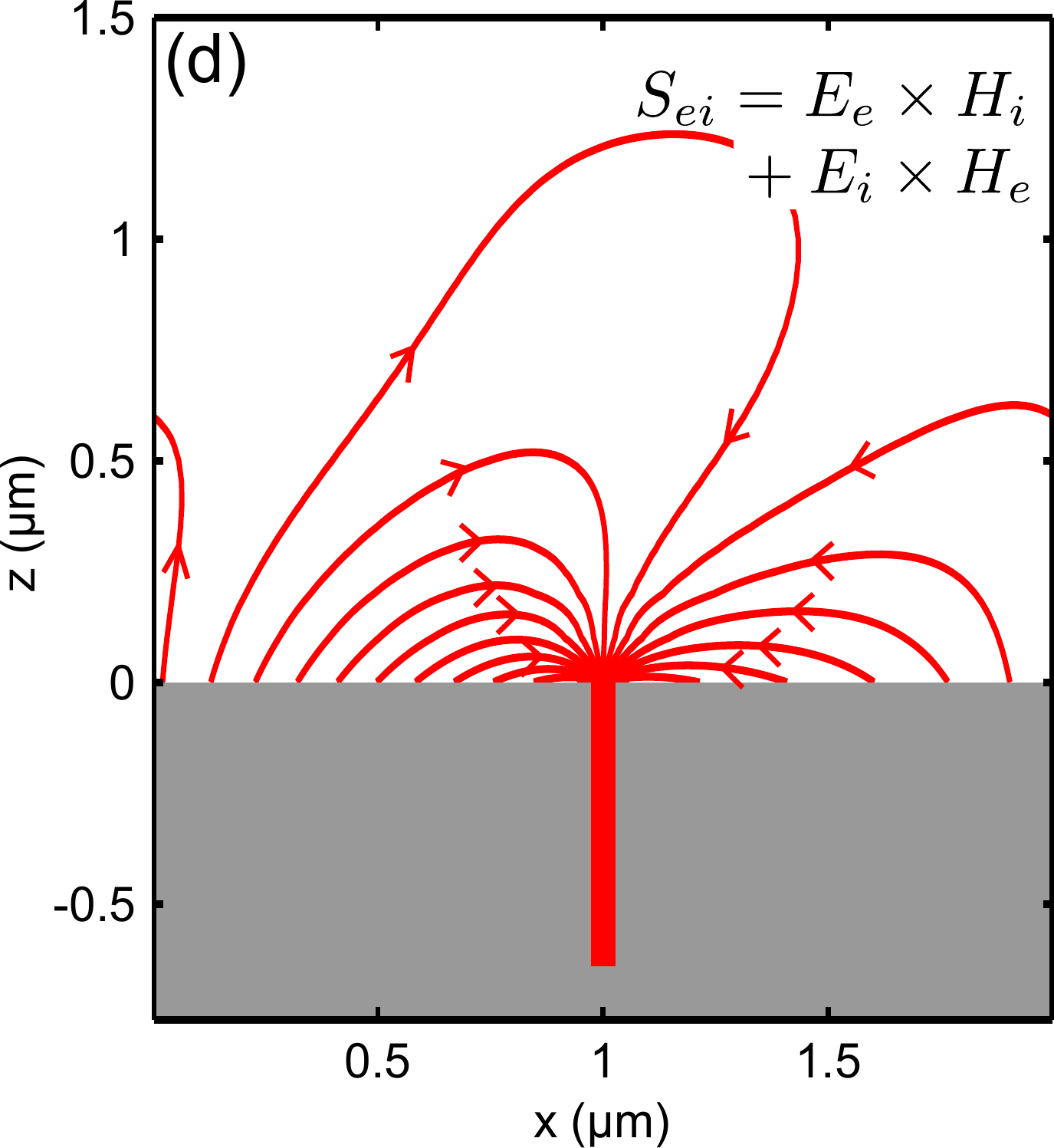}\vfill
  \includegraphics[width=0.45\linewidth]{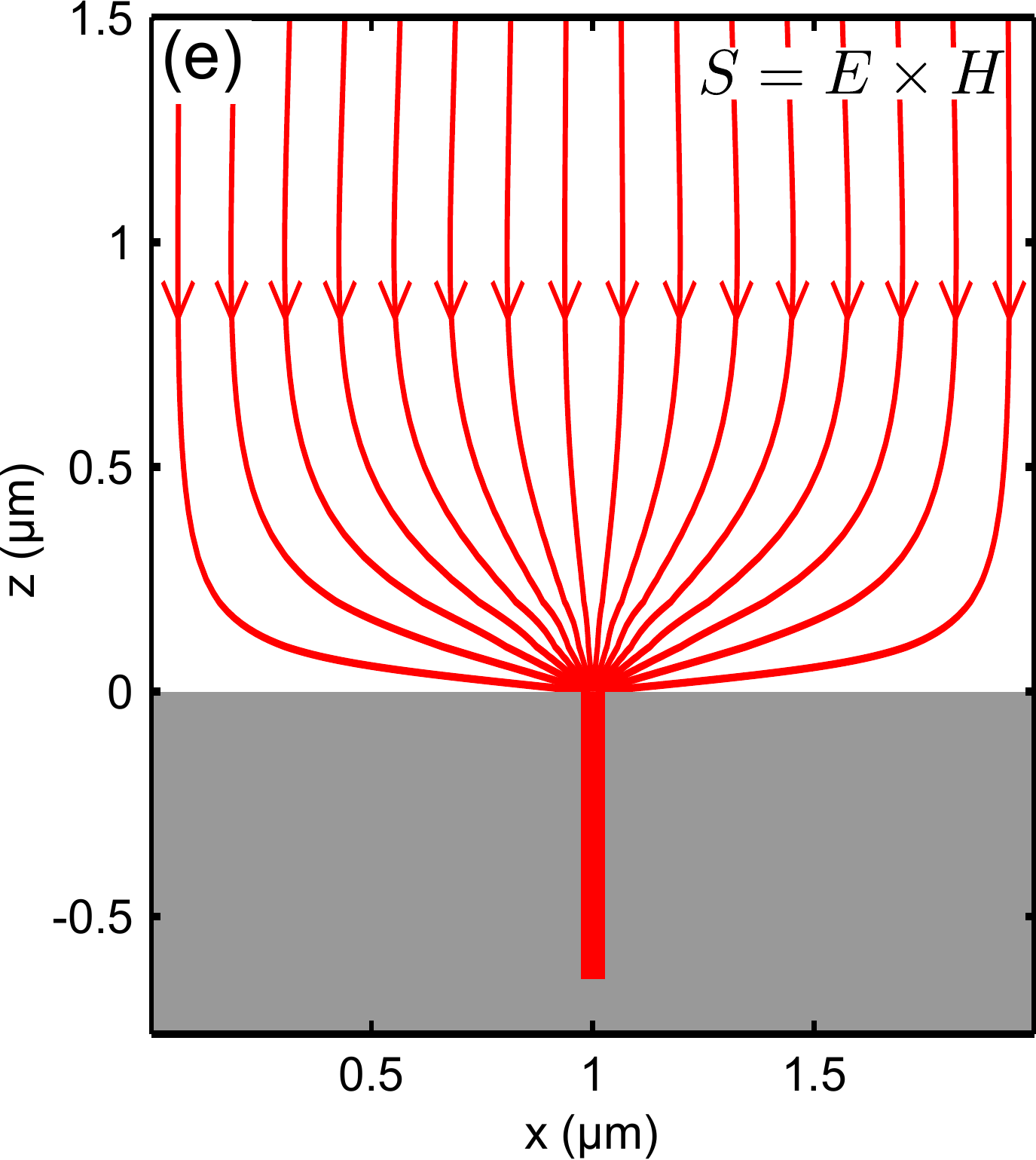}
  \includegraphics[width=0.45\linewidth]{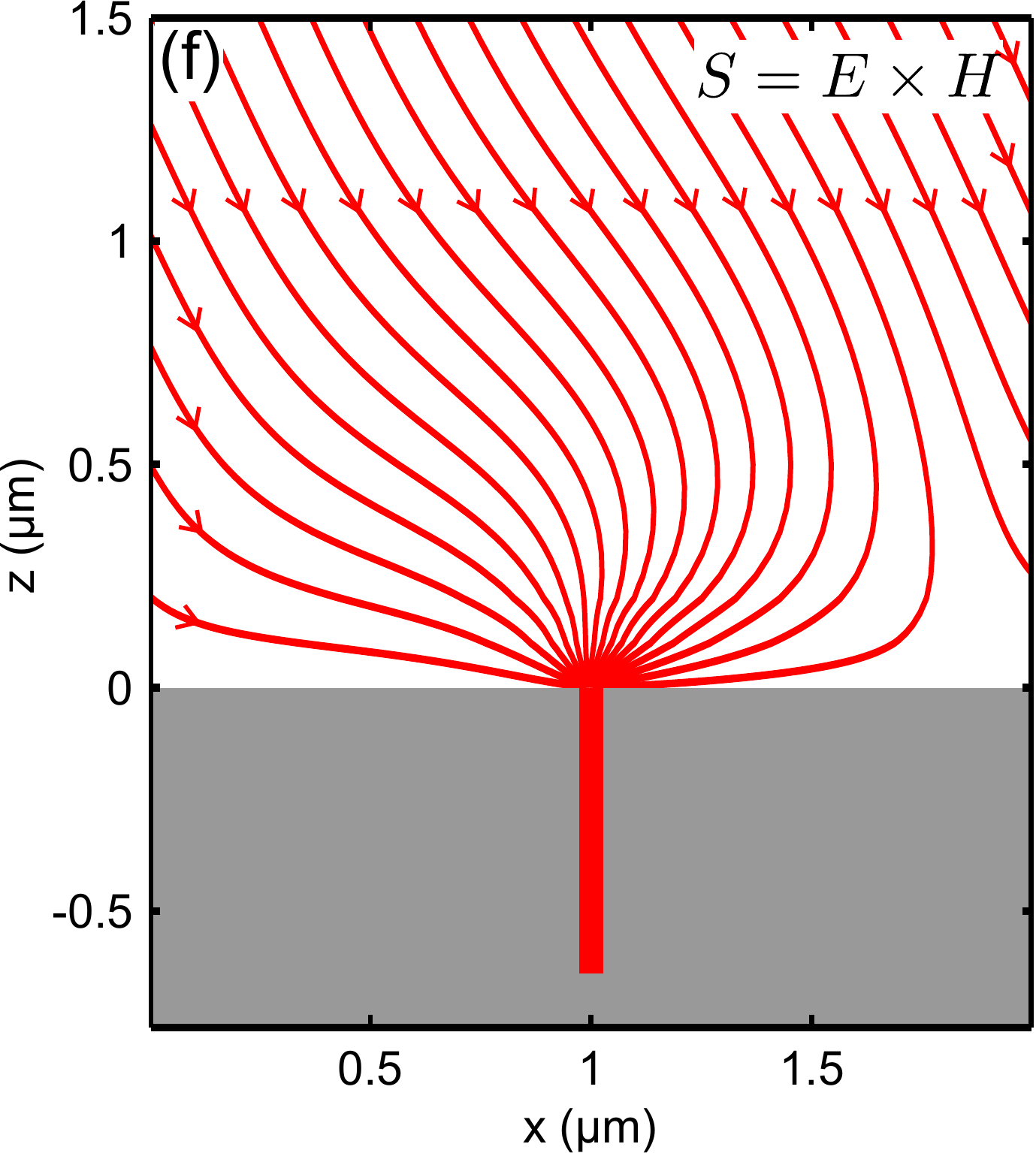}
  \caption{\label{fig:flux_poynt}  
    Poynting-vector streamlines in one period of the slit grating
    for two angles of incidence $\theta = 0\,^\circ$ (left column)
    and $\theta =30 \,^\circ$ (right column)
    at $\lambda= 4000 \, \mathrm{nm}$.
    Streamlines of the incident wave are shown on (a) and (b).
    Streamlines of the interference between
    the incident wave and the evanescent field
    are shown on (c) and (d).
    The energy flux of the evanescent waves is negligible
    in this structure for $\theta = 0\,^\circ$;
    refer to Fig. \ref{fig:flux_poynt_ev} (a)
    for an illustration at $\theta = 30\,^\circ$.
    Streamlines of the total Poynting vector are shown on
    (e) and (f).
    In both cases, the incident energy is funneled inside the groove
    where it is fully absorbed inside the metal.}
\end{figure}

The Poynting-vector streamlines for
$S_{i}$, $S_{ei}$ and $S$ are plotted at two angles of incidence
in Fig. (\ref{fig:flux_poynt}) so that
the flux of energy between two lines is constant.

At normal incidence, as expected for a propagative plane wave in air,
the lines for incident flux $S_{i}$ are equidistant
and in the propagation direction.
The MEI $S_{ei}$ lines
are coming from the surface and are converging on the groove.
On the metallic surface, they compensate for
the flux of the incident plane wave
and funnel it inside the groove.
By drawing lines perpendicular to the Poynting streamlines on
Fig. \ref{fig:flux_poynt} (c),
and taking in account that the predominant term in $S_{ei}$
is $E_e \times H_i$, one can deduce that the evanescent wave shape
is quasi-cylindrical \cite{ Natur-Liu2008mto,PhRvL-Yang2009ccb}.
The evanescent flux $S_{e}$ (not shown)
carries energy $1000$ times weaker and does not play an active role
in the funneling for this structure at normal incidence.
In Fig. \ref{fig:flux_poynt} (e),
the total flux of energy $S$ is shown to funnel into the groove
in the near-field region ($z \lesssim 500 \, \mathrm{nm}$).
Eventually all the incident flux is dissipated, mostly inside the groove.

\begin{figure}[ht]
  \includegraphics[width=0.45\linewidth]{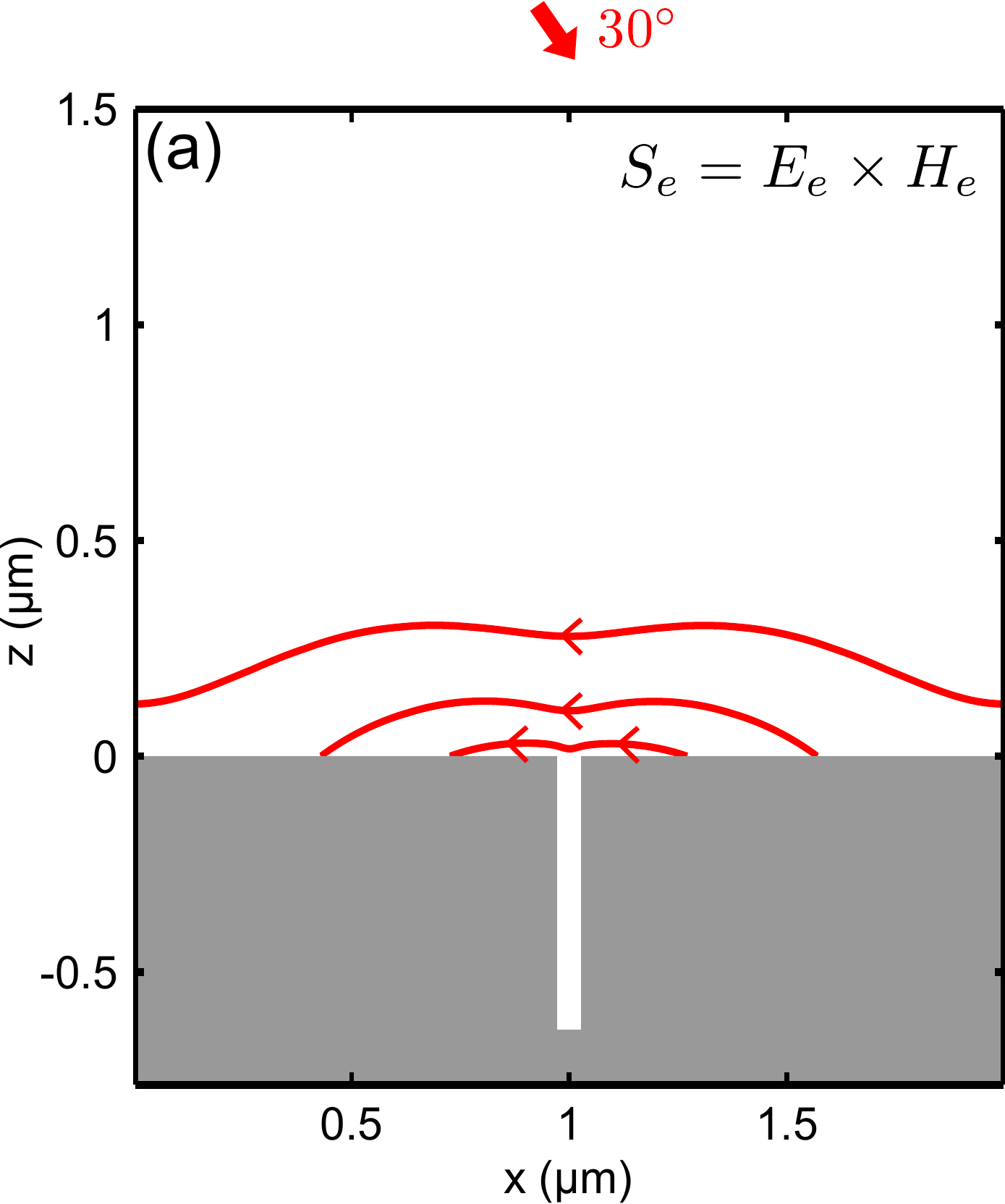}
  \includegraphics[width=0.45\linewidth]{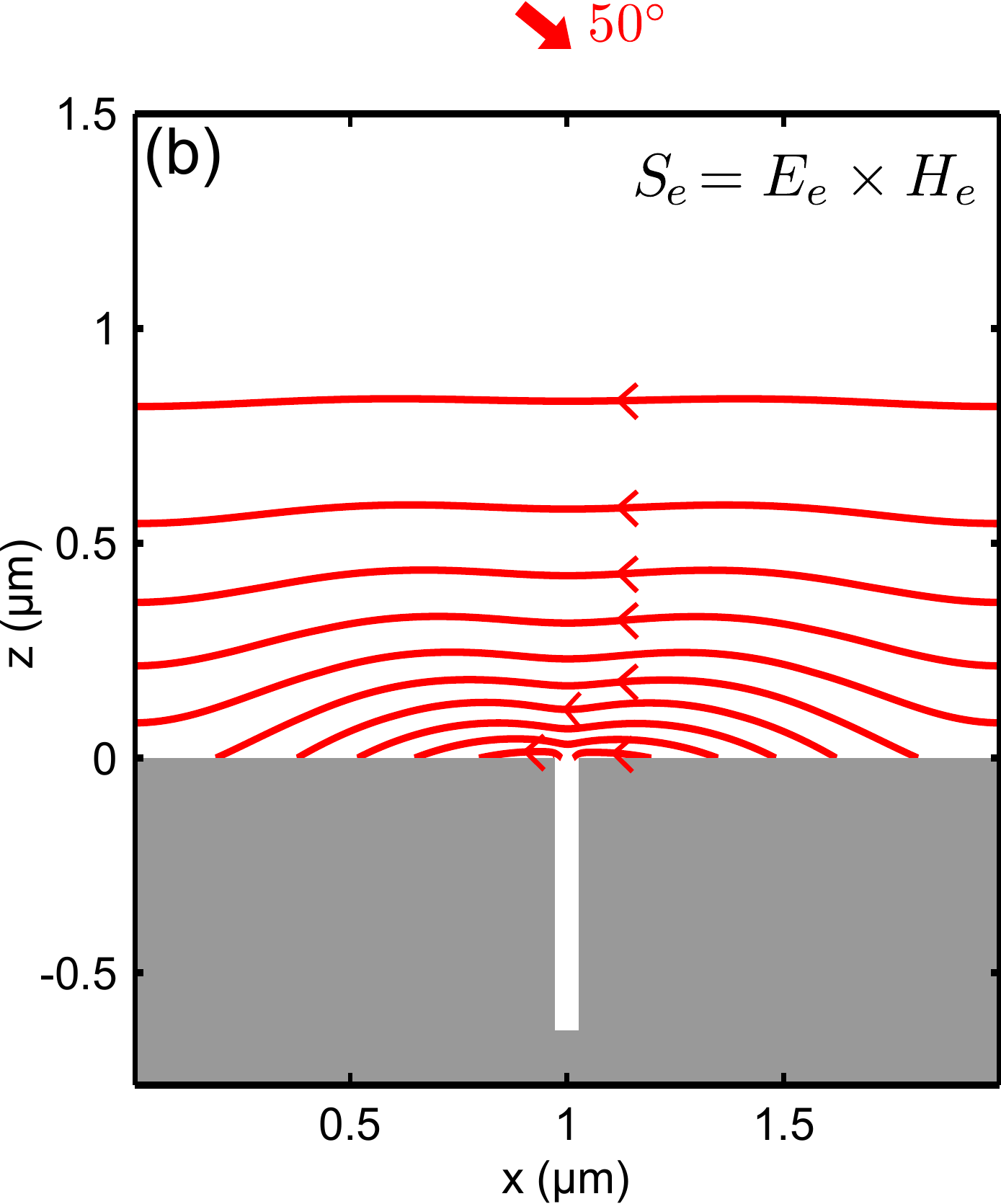}
  \caption{\label{fig:flux_poynt_ev} Poynting-vector streamlines
    of the evanescent field
    for two angles of incidence
    (a) $30 \,^\circ$ and (b) $50 \,^\circ$ at $\lambda= 4000 \, \mathrm{nm}$.
    $S_e$ does not play a role of funneling
    but redistributes the energy in the grating.}
\end{figure}

For an incidence of $30\,^\circ$,
the MEI
$S_{ei}$
is still funneling the energy towards the slits
(Fig. \ref{fig:flux_poynt} (d)).
But it no longer compensates for the incident flux
which lines are equidistant.
Indeed, in Fig. \ref{fig:flux_poynt} (d),
there are more lines going out from the metal surface
on the left (10 lines)
than on the right (6 lines).
Nonetheless, Fig. \ref{fig:flux_poynt} (f)
shows that the incident energy gets funneled into the groove
despite this assymetry. In fact, at oblique incidence, the evanescent field
carries an energy flux $S_{e}$
which is no longer negligible, as shown in Fig. \ref{fig:flux_poynt_ev}:
the energy is redirected from the right side of the groove to the left side.
This redistribution of the energy
compensates for the dissymmetry of $S_{ei}$
which appears in Fig. \ref{fig:flux_poynt} (d).
In conclusion $S_{e}$ plays no role
in the funneling, but helps to redistribute energy over the grating.
This incidentally invalidates the hypothetical role of plasmonic waves, 
which are evanescent waves, in the funneling 
mechanism. 
As a general comment, it is interesting to point out that the MEI process
is known to be responsible of the optical tunnelling
(frustrated total reflection)\cite{Book-Smith1997ait}.
Our study unveils its key role in a larger spectrum
of near-field energy transfer phenomena.

\begin{figure}[ht]
  \includegraphics[width=0.45\linewidth]{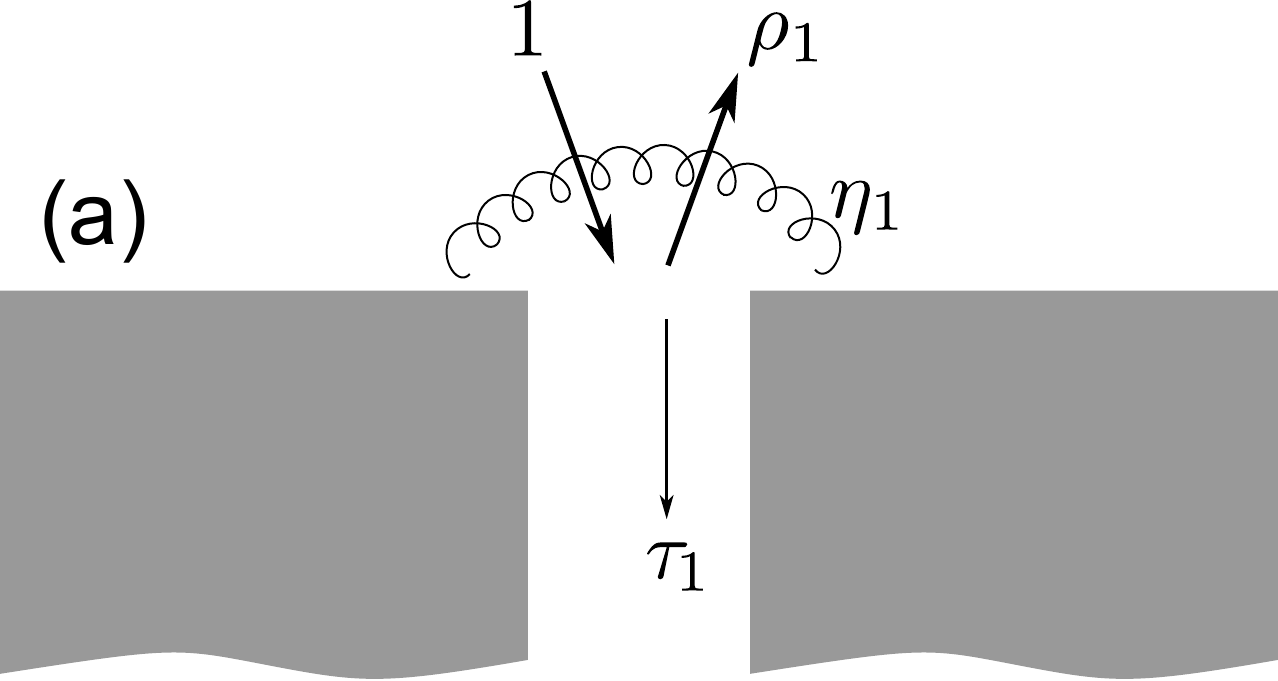}
  \includegraphics[width=0.45\linewidth]{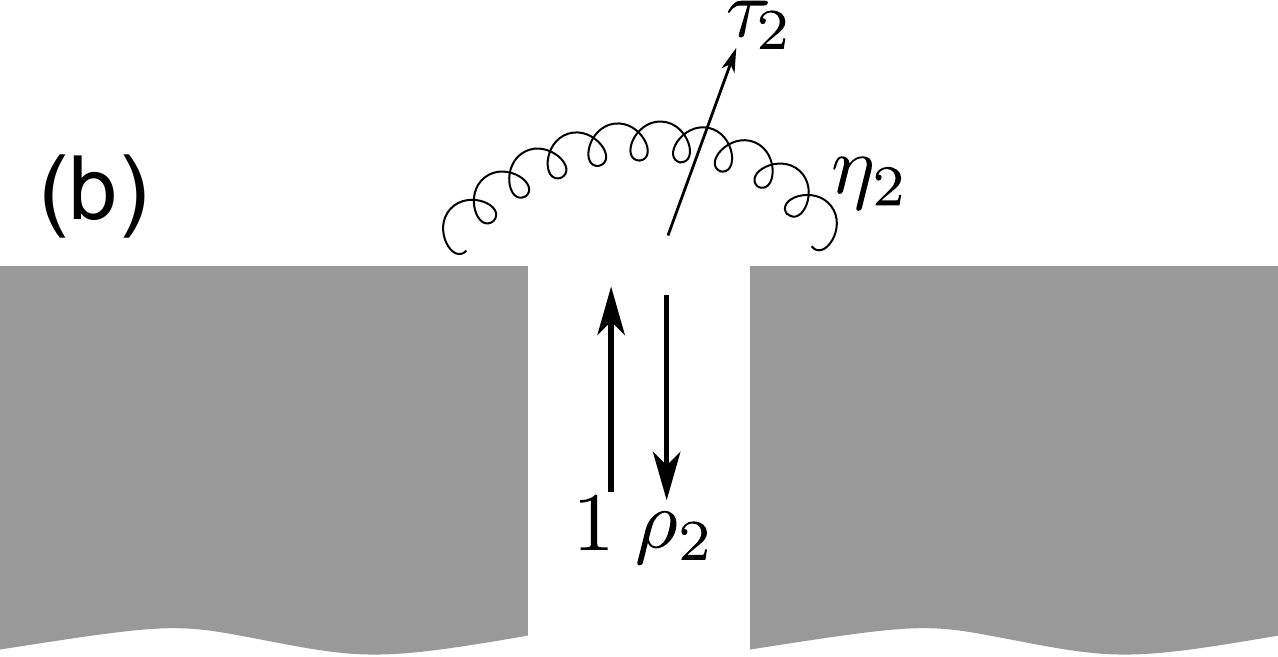}
  \vfill\vspace{1mm}
  \includegraphics[width=0.45\linewidth]{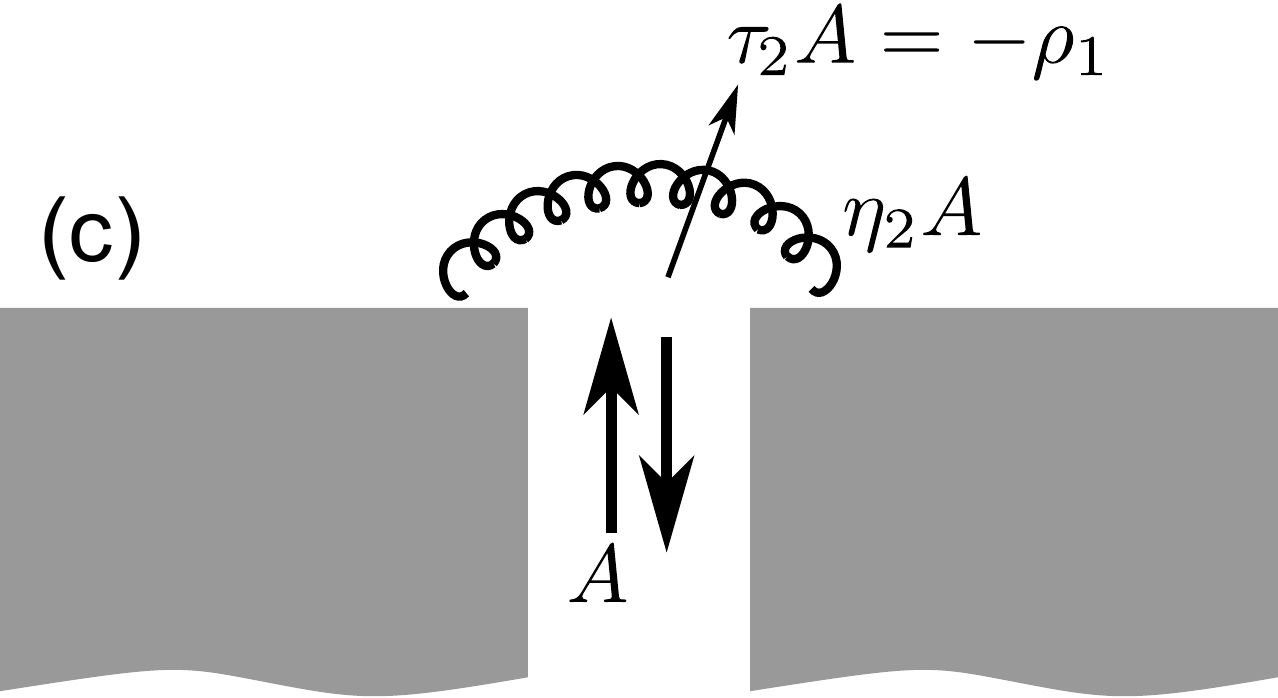}
  \includegraphics[width=0.45\linewidth]{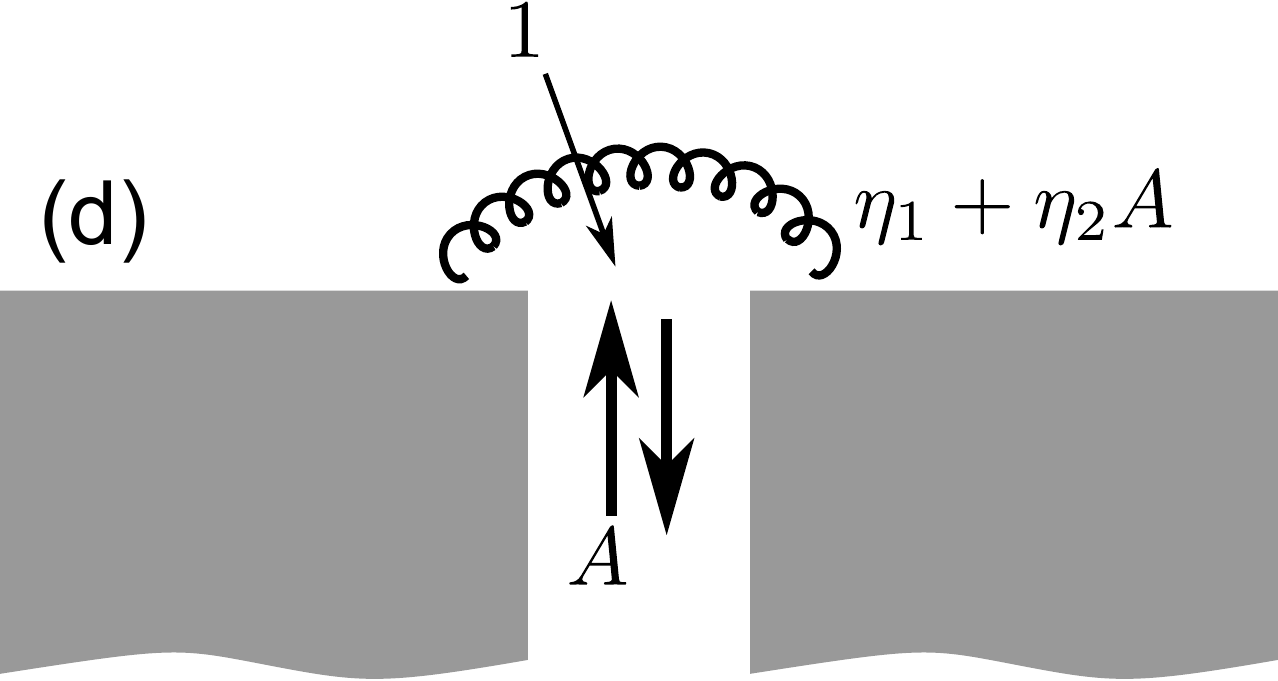}
  \caption{\label{fig:fp} Isolated single interface analysis
    of the metallic grating optimized to absorb all the incident light.
    (a) Unit plane wave from air: $\rho_{1}$ is the amplitude of the reflected
    plane wave, $\eta_{1}$ is the vector of evanescent field amplitudes. 
    (b) Unit modal wave from bottom of grooves: $\tau_{2}$ is the amplitude
    of the plane wave escaping in the air,
    $\eta_{2}$ is the vector of evanescent field amplitudes.
    (c) Same as (b), but with the modal wave having
    the amplitude $A$.
    (d) Superposition of (a) and (c), showing the field amplitudes
    in the grating
    excitated by an unit plane wave: the reflectivity is null
    and the evanescent field is dominated by the term
    escaping from the resonator.
    }
\end{figure}

We now aim to describe the origin of the evanescent
field involved in the funneling process. The subwavelength
groove behaves as a Fabry-Perot resonator.
As in ref. \cite{PhRvB-Lalanne2003paf}, we consider the isolated single interface
in two configurations: one where the incident field is a unit-amplitude
plane wave in air; the other where the incident field is a wave coming from
the bottom of the groove, and is unit-amplitude at the interface.
In the first case (Fig. \ref{fig:fp} (a))
the reflected wave has an amplitude $\rho_{1}$ and an 
evanescent field $\eta_{1}$. Due to the low aperture ratio $w/d$, we get
$|\rho_{1}| \lesssim 1$ and $|\eta_{1}| \ll 1$. In the
second case  (Fig. \ref{fig:fp} (b)) the transmitted wave (which
corresponds to the reflected wave above)
has an amplitude $\tau_{2}$ and the evanescent field in the air is written
$\eta_{2}$. If a unit-amplitude wave is defined as
taking the value $H_y = 1$ at the center of a groove entrance, then
computation shows that evanescent amplitudes $\eta_1$ and $\eta_2$
are nearly equal. 
Due to the low aperture ratio $w/d$, we get $|\tau_{2}| \ll 1$.

At the resonance, the wave coming from
the bottom of the groove has an
amplitude $A = -\tau_{2}^{-1} \rho_{1}$
so that all the amplitudes at sake in Fig. \ref{fig:fp} (b) are
multiplied by the factor $A$, which leads to Fig. \ref{fig:fp} (c).
The response of the grating excited by a unit-amplitude plane
wave at this resonance is thus given by the superposition of amplitudes
of Fig. \ref{fig:fp} (a) and (c), shown on  Fig. \ref{fig:fp} (d).
This leads to the expected
null-amplitude reflected wave. Moreover, the resulting
evanescent field is expressed as $\eta_{1} + \eta_{2}A \simeq \eta_{2}A$ because
$|\eta_{1}| \simeq |\eta_{2}|$ and $A \gg 1$
(e.g., for the grating described in Fig. \ref{fig:reflec}, one computes
$|A| \simeq 11$).
To summarize, the wave built inside the grooves escapes in the air as both
a propagating plane wave $\tau_{2} A$ and an evanescent field $\eta_{2} A$ (see Fig. \ref{fig:fp} (c)).
The propagating plane wave interferes destructively
with the directly reflected wave ($\tau_{2} A + \rho_1 = 0$)
leading to the null reflection,
and the evanescent field interferes
with the incoming plane wave to funnel the energy into the groove.
The two effects are of course not independent: we have no reflection
because the energy is nearly fully collected in the groove.
As a generalization path, it is interesting to note that this analysis still
holds whatever the structure etched behind the aperture on the metal surface
(rectangular slits, or more complex shapes such as in Ref.\cite{NaPho-Teperik2008oai}).

Finally, we want to highlight that the key role played by
the MEI of propagative waves with the evanescent field in
the funneling mechanism is not limited to resonant structures.
Let us thus consider the non-resonant case of a grating
made of infinitely deep grooves, i.e. the simple isolated
interface illustrated in Fig. \ref{fig:fp}(a),
with a reflected wave of amplitude
$\rho_1$. For an incidence angle of 30$^{\circ}$, we
get $|\rho_1|^2=0.83$: about 17\% of the incident
energy enters the grooves. This value is much
larger than the aperture ratio
$w/d=2.8 \%$, thus there appears to be funneling.
Now, if we compare this to the situation of Fig. \ref{fig:fp}(d),
the evanescent field of Fig. \ref{fig:fp}(a)
is lowered by a factor $|\eta_1 + \eta_2 A|/|\eta_1| \simeq |A| \simeq 11$.
We therefore compute that the interference of the incident wave
with the evanescent field gives a funneling of about $1/|A| \simeq 9\%$.
The missing 8\% stems from the interference of the evanescent field
with the reflected wave $\rho_1$, which
paradoxically contributes to the funneling of energy inside the grooves.
The key
to the paradox lies here: firstly, the main contribution
to the interference is $E_e \times H_r$ ($E_r \times H_e$ is much smaller,
at least at the interface level) and, secondly, the magnetic field $H_r$
has the sign of $H_i$ due to the metallic reflection. 

In conclusion, we have unveiled the funneling
mechanism of incident light
in very narrow grooves etched on a metal surface. It originates
from the magneto-electric interference between the incident wave and the
evanescent field, in both resonant and non resonant situations.
Furthermore, this result has been generalized to any sub-wavelength aperture
etched on a metal surface (whatever the structure behind it)
thanks to a single interface analysis. 
In the resonant case, the evanescent field escaping
from the apertures can lead to the full harvesting of
incident photons for a broad range of incidences.
From a practical point of view, this approach
opens a new avenue for the design of electromagnetic
resonant antennas, based on the tailoring of the escaping evanescent field.

Eventually, we have shown that evanescent waves propagating along
the interface do not carry any energy through the apertures.
This clearly demonstrates that the funneling 
is not mediated by plasmon waves at the surface.

\begin{acknowledgments}
This work was partially supported by the ANTARES Carnot project.
\end{acknowledgments}

\bibliography{EF_20110607}
\setcounter{figure}{0}
\renewcommand\thefigure{S\arabic{figure}} 

\section{Supplemental material}

Supplementary information for the article \emph{Light funneling mechanism explained by magneto-electric interference} (MEI). It includes the following items:
\begin{itemize}
\item Broadband extraordinary transmission analysis by MEI.
\item Experimental results on groove gratings described in the article.
\item Field maps.
\item Scaling properties.
\end{itemize}

\subsection{Broadband extraordinary transmission analysis by MEI} 

\begin{figure}[h!]
  \includegraphics[width=1.0\linewidth]{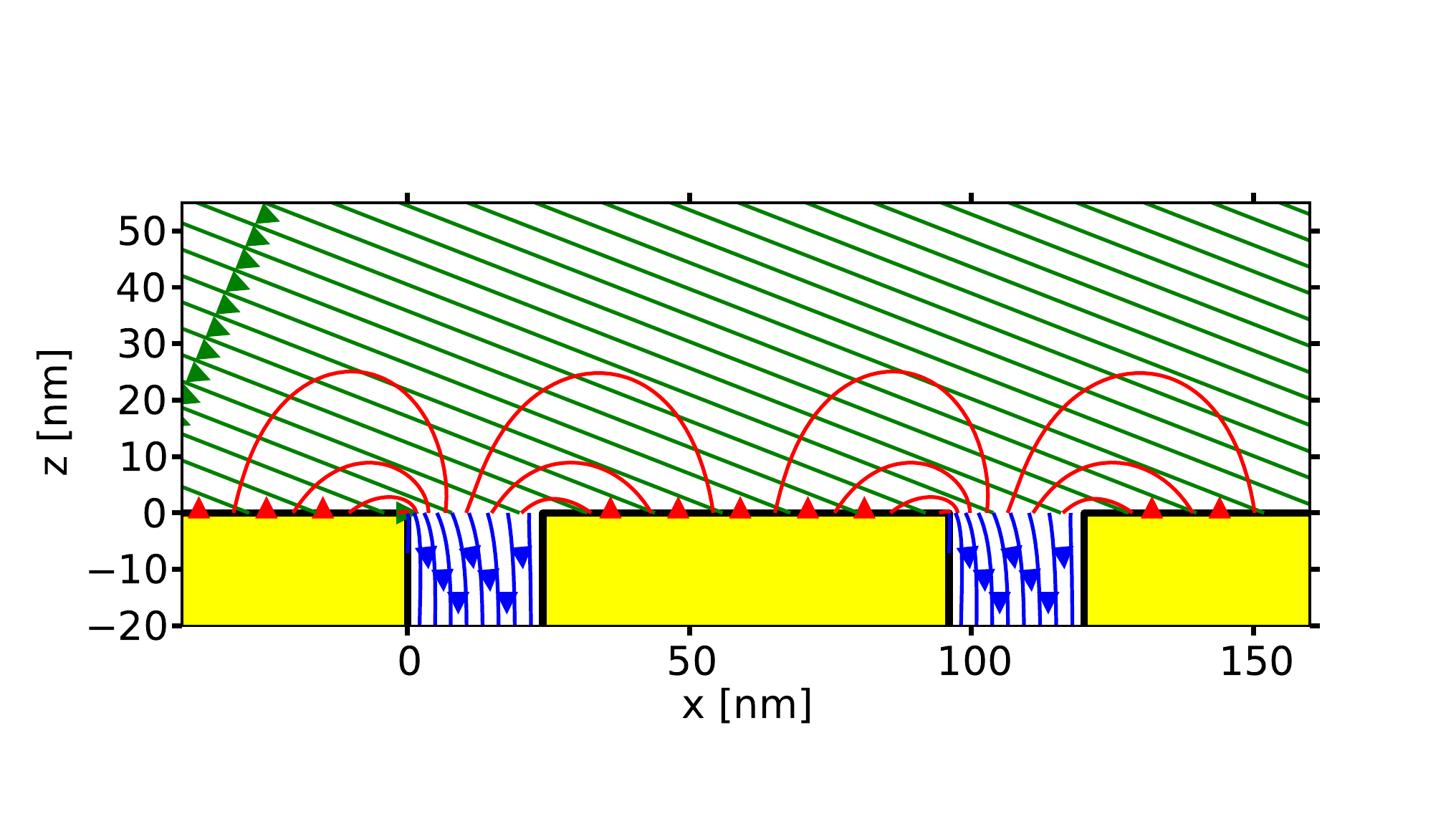}
  \caption{\label{fig:brewster_mei}Decomposition of incident energy flux of figure \ref{fig:brewster_global} in two terms: the incident energy flux (straight green lines), and the magneto-electric interference (MEI, curved red lines). All other terms are null. Hence, the total funneling is fully explained by MEI energy collection from the surface to the slits.}
\end{figure}

\begin{figure}[h!]
  \includegraphics[width=0.9\linewidth]{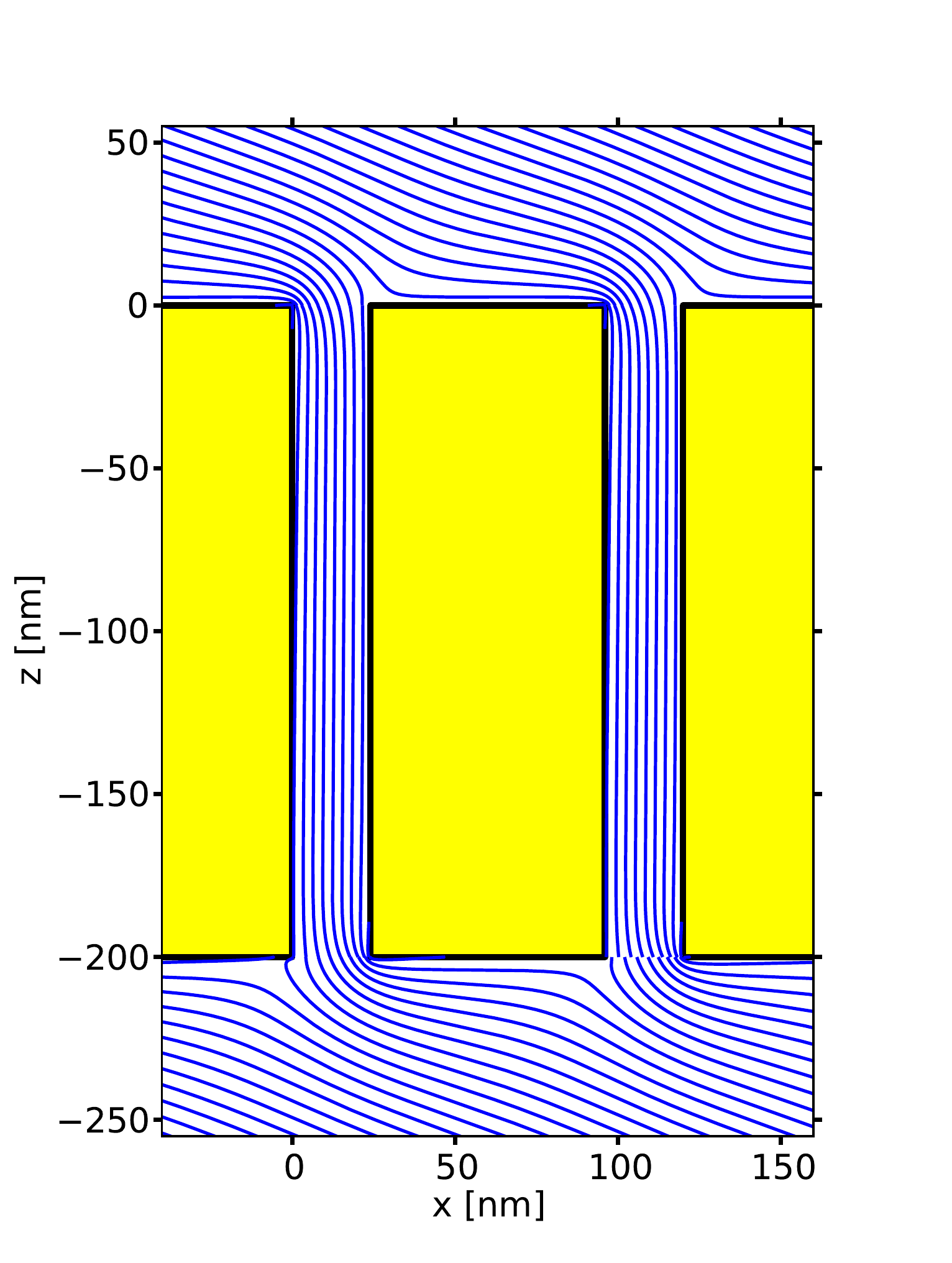}
  \caption{\label{fig:brewster_global}
    Poynting vector streamlines in the broadband extraordinary transmission
with parameters of Ref. \cite{PhRvL-Alu2011pba} (wavelength $\lambda$ = 759 nm, angle of incidence $\theta = 68.60^\circ$, metal permittivity
$\epsilon_m = -24.7 + 1.44j$, slit width $w$ = 24 nm, period $d$ = 96 nm, height $h$ = 200 nm).
Incident wave arrive from the top left.
There is strictly no resonance inside the slit and no reflexion at all.
The energy flux between two lines is constant, equal to 1/8 of the incident energy flux on one period.
Lines need to be cut around a metal bloc (here at bottom of the right slit) because of the losses.}
\end{figure}

The total funneling in the case of
broadband extraordinary transmission described by Al\`u \emph{et al.}
\cite{PhRvL-Alu2011pba} is illustrated on figure
\ref{fig:brewster_global} and fully explained by MEI as shown on figure
\ref{fig:brewster_mei}. In constrast with fig. 3, the evanescent
fields (amongst these evanescent fields are the surface plasmon contribution),
carry here strictly no energy. As a result, in both normal and Brewster
incidences, MEI alone redirects all the incident energy towards the slits.

\subsection{Experimental results}

\begin{figure}[ht]
  \includegraphics[width=0.9\linewidth]{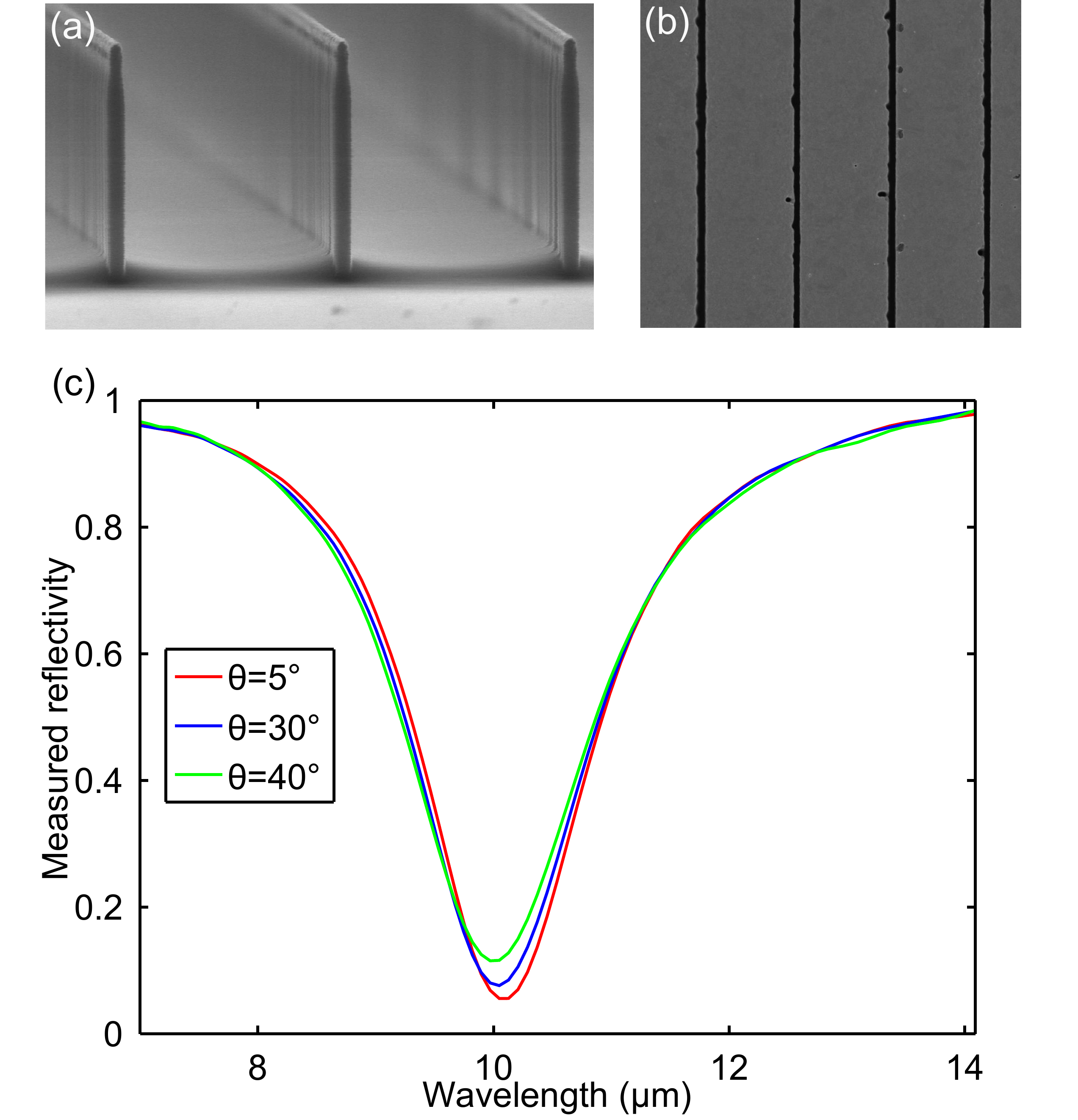}
  \includegraphics[width=0.9\linewidth]{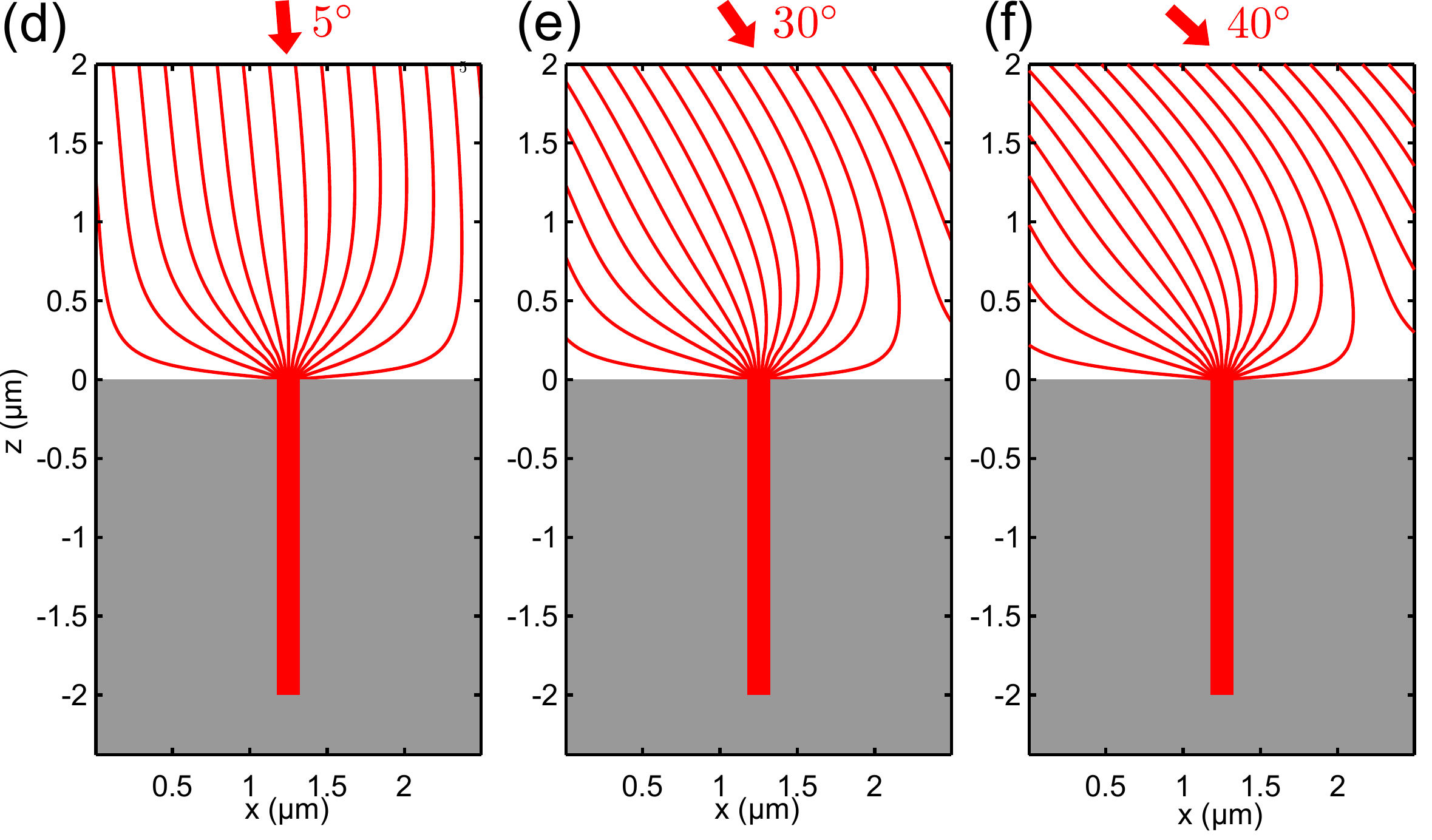}
  \caption{\label{fig:groove_experiments}  
    (a) SEM image of the grating of GaAs walls which is
    used as a mold. (b) SEM image of the grating of 
    gold slits ($w=150$nm, $h=2\mu \mathrm{m}$ and $d=2.5\mu \mathrm{m}$) obtained
    after an electroplating on the GaAs mold. (c) Reflectivity curves of the
    sample for various incident angles $\theta=5^\circ$, $\theta=30^\circ$ and
    $\theta=40^\circ$. The light is TM-polarized. The resonance peak wavelength 
    is nearly independent of the incidence angle and leads to an almost total
    absorption.
    Poynting vector streamlines at the resonance peak $\lambda=10.05 \mu \mathrm{m}$
    for an incidence of (d) $\theta=5^\circ$, (e) $\lambda=30^\circ$ and
    (f)  $\theta=40^\circ$. The reflected energy is neglected.}
\end{figure}

The theoretical computations presented in the article are confirmed by experimental results.
To fabricate such a grating of high aspect ratio metallic grooves, we have developped
a mold cast technique based on a gold electroplating. The mold in GaAs is fabricated 
trhough inductively coupled plasma reactive ion etching (ICP-RIE). A scanning
electron microscope (SEM) image of the GaAs walls is shown in figure 
\ref{fig:groove_experiments}a, they are $2\mu \mathrm{m}$ high and $150$nm wide
in a period $d=2.5\mu \mathrm{m}$. A gold electroplating is then done on the 
GaAs mold which is eventually chemically removed, leading to the grating of 
gold grooves (see SEM image in figure \ref{fig:groove_experiments}b).
Measured reflectivity spectra of a TM incident wave at various angles ($\theta=5^\circ$,
$\theta =30^\circ$ and $\theta=40^\circ$) are shown on figure \ref{fig:groove_experiments}c.
They were obtained using a Fourier transform infrared spectrometer and an home-made
achromatic optical system \cite{billaudeau2008angle}. As predicted, there is an
almost total extinction of light at $\lambda=10.05\mu \mathrm{m}$ which is 
nearly independent of the incidence angle. The measurements are in 
excellent agreement with the computed spectra. 
The corresponding funneling behavior of the poynting energy flux are shown for 
these three incidence angles in figure \ref{fig:groove_experiments}d-f. The 
reflected energy is neglected in these three cases. 

\subsection{Field maps}

\begin{figure}[!ht]
  \includegraphics[width=0.9\linewidth]{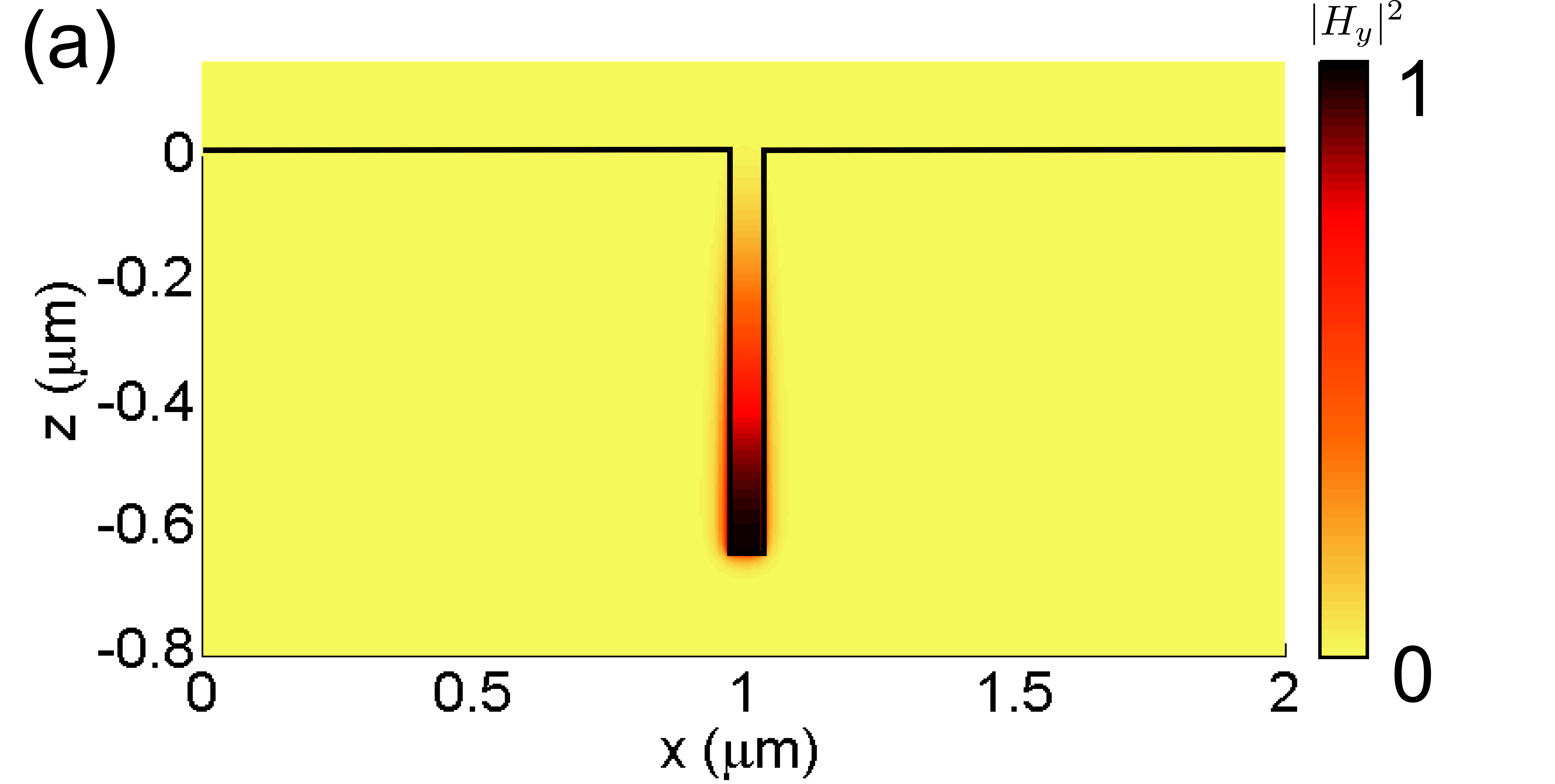}
  \includegraphics[width=0.85\linewidth]{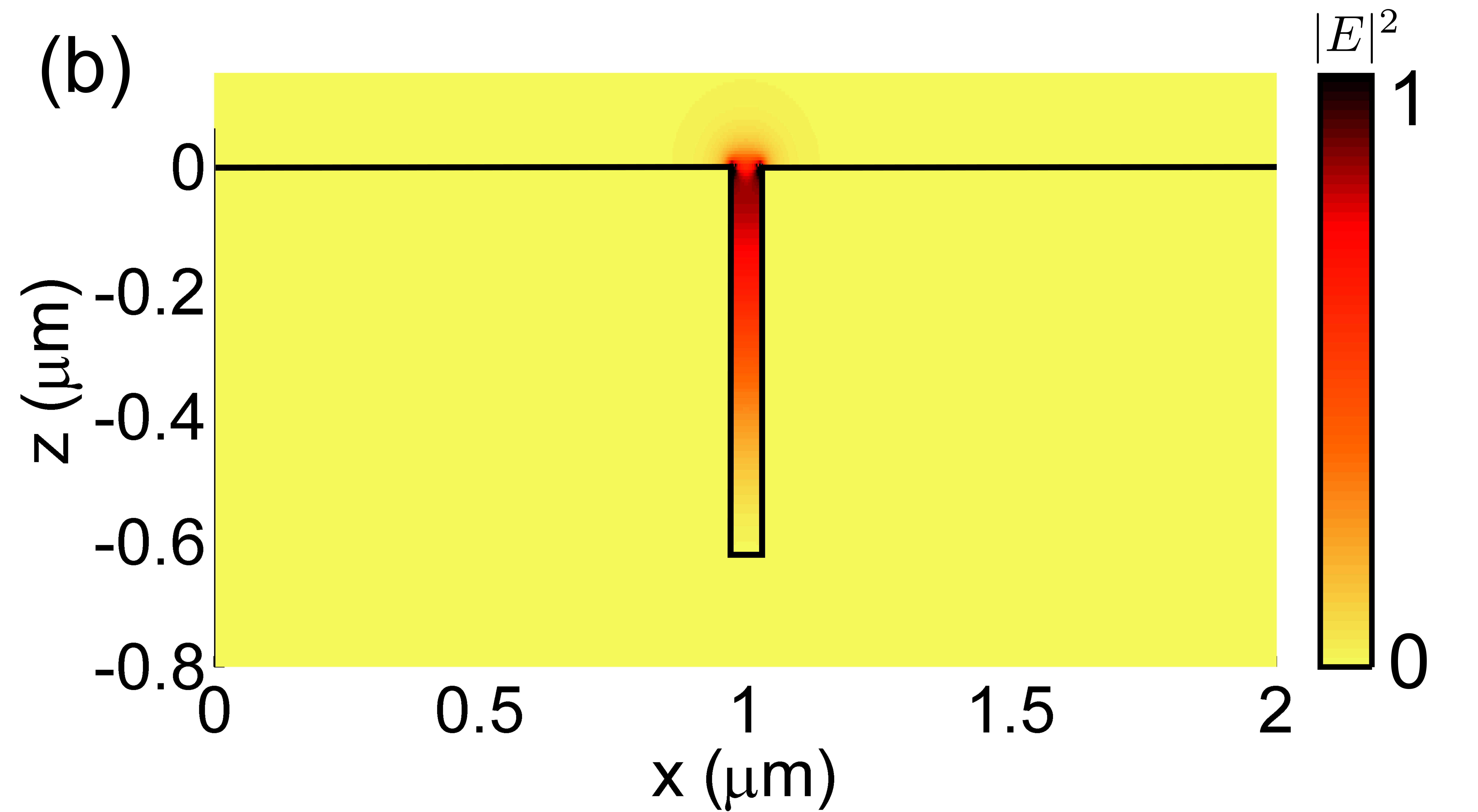}
  \caption{\label{fig:map_field_groove}  
    Magnetic and electric fields ((a): $|H_y|^2$, (b): $|E|^2$) in the groove
    of width $w=56$nm, depth $h=640nm$ and period $d=2\mu \mathrm{m}$ at the resonance peak
  $\lambda _p=4 \mu \mathrm{m}$ at normal incidence.}
\end{figure}

The magnetic and electric fields maps are shown in figure 
\ref{fig:map_field_groove} a-b for the groove of width $w=56$nm, depth $h=640nm$ 
and period $d=2\mu \mathrm{m}$ at the resonance peak wavelength
  $\lambda _p=4 \mu \mathrm{m}$. The incident wave is at normal incidence and 
  in TM polarization.
  The magnetic field $H_y$ is concentrated at the bottom of the groove ($z=-640$nm) 
  while the
  electric field is concentrated at the aperture of the groove ($z=0$nm).

\subsection{Scaling properties}

\begin{figure}[ht]
  \includegraphics[width=0.9\linewidth]{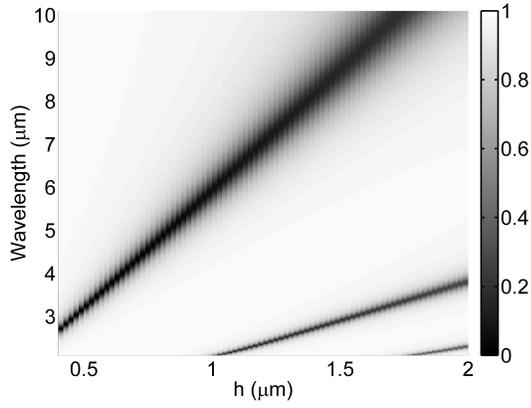}
  \caption{\label{fig:groove_as_h}  
    Reflectivity map as a function of the groove's depth. Other parameters
    are kept constant (width $w=56$nm and period $d=2\mu \mathrm{m}$). 
    The incident wave is TM-polarized and coming at normal incidence.}
\end{figure}

The dependence of the resonance peak as a function of the groove's depth $h$
 is studied in the reflectivity map plotted on figure \ref{fig:groove_as_h}. 
The width of the groove $w=56$nm and
 the period of the grating $d=2\mu \mathrm{m}$ are kept constant. 
 The incident wave is TM-polarized and has an incidence of $\theta=0 ^\circ$. 
 There are three lines with total absorption of light appears. They correspond 
 to the first three cavity modes inside the grooves. These resonances position
 as a function of the groove's depth is given by the law
 $\lambda _m=4h n_{\mathrm{eff}}/(2m-1)+\phi$ where $m$ is an integer standing for the 
 resonance order,
 $n_{\mathrm{eff}}$ is the effective index of the guided mode inside the groove
 and $\phi$ is the wavelength shift induced by the reflection of the guided mode
 on the grating surface interface.   
At the first order, the effective index takes a very simple form: 
\begin{equation}
n_{\mathrm{eff}}=1+\delta /w 
\label{eq:effective-index}
\end{equation}
where $\delta$ is the skin depth of the metal
\cite{OExpr-Collin2007wmi}. 

\begin{figure}[!h]
  \includegraphics[width=0.9\linewidth]{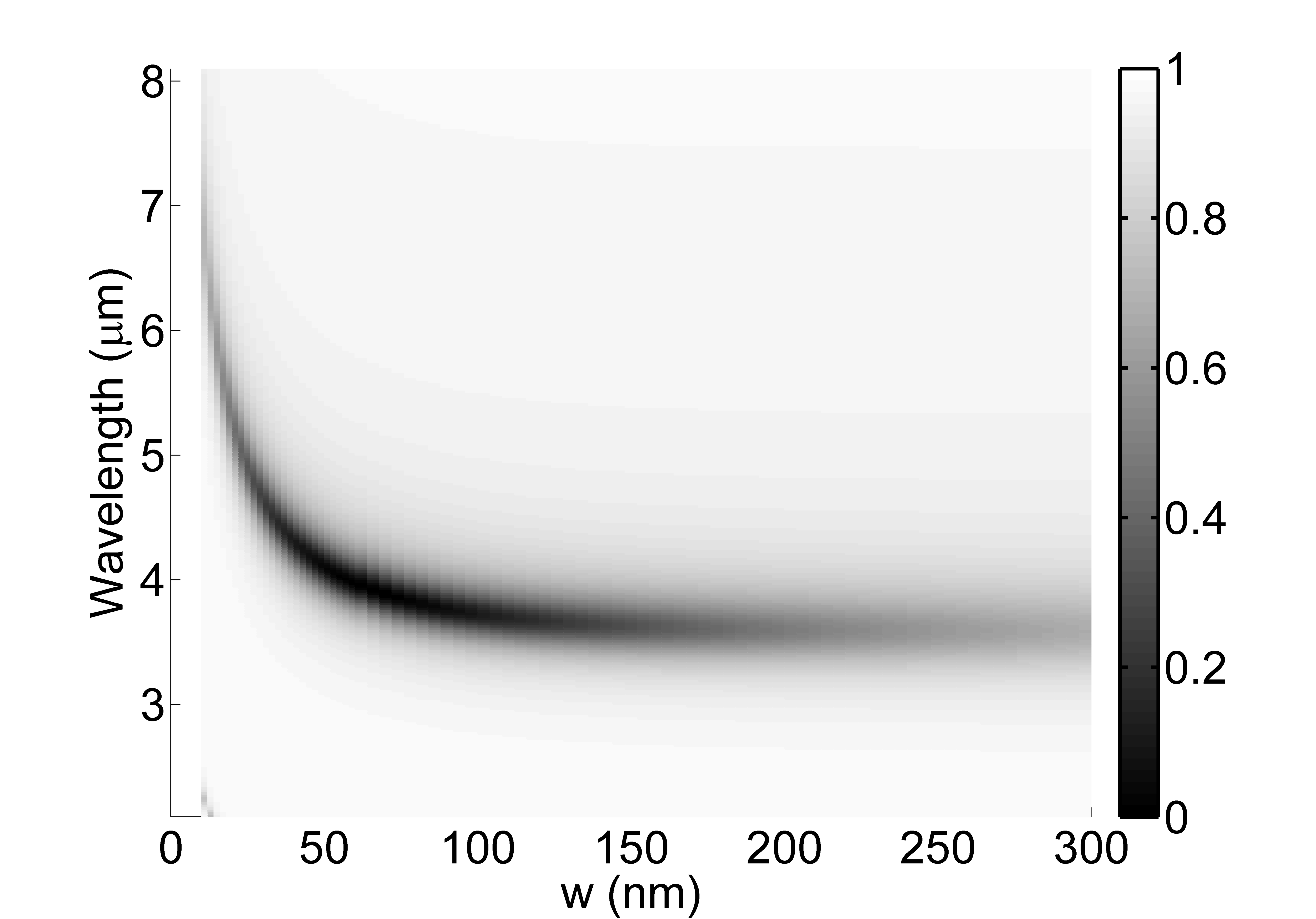}
  \caption{\label{fig:groove_as_w}  
    Reflectivity map as a function of the groove's width. Other parameters
    are kept constant (depth $h=640$nm and period $d=2\mu \mathrm{m}$). 
    The incident wave is TM-polarized and coming at normal incidence.}
\end{figure}
The dependence of the resonance peak as a function of the groove's width is 
studied in the reflectivity map plotted on figure \ref{fig:groove_as_w}.
The depth of the groove $h=640$nm and the period of the grating $d=2\mu \mathrm{m}$
are kept constant. 

The inverse dependance of the resonance peak with the groove's width is coherent
with the previous law given in equation \ref{eq:effective-index}.
\begin{figure}[!h]
  \includegraphics[width=0.9\linewidth]{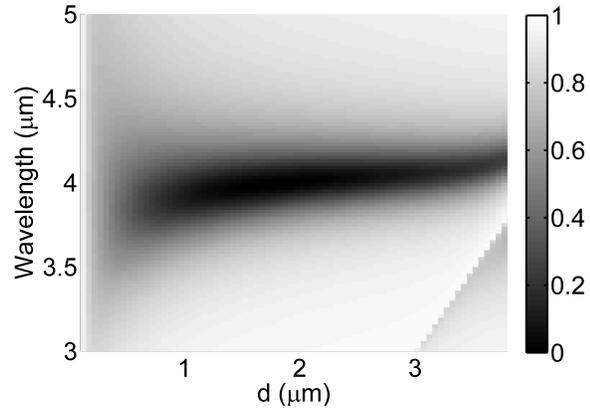}
  \caption{\label{fig:groove_as_d}  
    Reflectivity map as a function of the grating's period. The groove's width and 
    depth are kept constant ($w=56$nm and $h=640$nm). 
    The incident wave is TM-polarized and coming at normal incidence.
}
\end{figure}
The dependence of the resonance peak as a function of the grating's period is 
studied in the reflectivity map plotted on figure \ref{fig:groove_as_d}.
The depth $h=640$nm and width $w=56$nm of the groove are kept constant, which
set the resonance peak at $\lambda _p \simeq 4\mu \mathrm{m}$ for subwavelength
period. The reflectivity map confirms this weak influence of the period value on
the resonance peak.

\end{document}